\documentclass[11pt]{article}

\usepackage{hydrosophy}

\usepackage{tabularx}
\usepackage{multirow}
\usepackage{colortbl}
\definecolor{headergray}{gray}{0.92}
\definecolor{metricblue}{RGB}{220,230,242}

\usepackage{algorithm}
\usepackage{algorithmic}

\usepackage{marvosym}
\usepackage{accsupp}
\newcommand{\corrmark}{\BeginAccSupp{method=hex,unicode,ActualText=2709}\Letter\EndAccSupp{}}
\newcommand{\githubicon}{\raisebox{-1.5pt}{\includegraphics[height=1.05em]{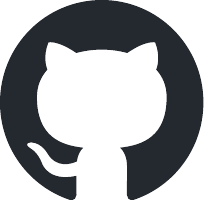}}}

\title{ToxiEval-ZKP: A Structure-Private Verification Framework for Molecular Toxicity Repair Tasks}
\shorttitle{ToxiEval-ZKP}

\author[1,2,3]{Fei Lin}
\author[1,2]{Tengchao Zhang}
\author[1,3]{Ziyang Gong}
\author[2,4,\corrmark]{Fei-Yue Wang}

\affiliation[1]{HydroSophy}
\affiliation[2]{Macau University of Science and Technology}
\affiliation[3]{Shanghai Jiao Tong University}
\affiliation[4]{Institute of Automation, Chinese Academy of Sciences}

\authornote[\corrmark]{Corresponding author.}

\metadata[\githubicon\ Code]{\url{https://github.com/HydroSophy/ToxiEval-ZKP}}

\begin{document}

\maketitle

\begin{abstract}
In recent years, generative artificial intelligence (GenAI) has demonstrated remarkable capabilities in high-stakes domains such as molecular science. However, challenges related to the verifiability and structural privacy of its outputs remain largely unresolved. This paper focuses on the task of molecular toxicity repair. It proposes a structure-private verification framework—\textbf{ToxiEval-ZKP}—which, for the first time, introduces zero-knowledge proof (ZKP) mechanisms into the evaluation process of this task. The system enables model developers to demonstrate to external verifiers that the generated molecules meet multidimensional toxicity repair criteria, without revealing the molecular structures themselves. To this end, we design a general-purpose circuit compatible with both classification and regression tasks, incorporating evaluation logic, Poseidon-based commitment hashing, and a \texttt{nullifier}-based replay prevention mechanism to build a complete end-to-end ZK verification system. Experimental results demonstrate that ToxiEval-ZKP facilitates adequate validation under complete structural invisibility, offering strong circuit efficiency, security, and adaptability, thereby opening up a novel paradigm for trustworthy evaluation in generative scientific tasks.
\end{abstract}

\section{Introduction}

In recent years, generative artificial intelligence (GenAI) has demonstrated tremendous potential in scientific domains. Generation technologies represented by large language models (LLMs), multimodal large language models (MLLMs), and diffusion models (DMs) have been widely applied to key tasks such as protein structure design, molecular property optimization, and novel drug candidate generation~\cite{lin2025autonomous, yim2024diffusion}. However, as GenAI systems increasingly engage in high-risk and high-value scientific tasks, issues of verifiability and structural privacy of their outputs have become more prominent.

A typical challenge lies in the lack of cryptographically verifiable mechanisms for model-generated results, making it difficult to credibly prove their scientific validity and regulatory compliance to external parties. For example, in the drug discovery process, a researcher may generate a molecule with significantly reduced toxicity. Yet without disclosing its specific structure, it is hard to convincingly demonstrate its safety and efficacy to regulators, collaborators, or academic reviewers. Moreover, molecular structures often carry core intellectual property, and disclosing them may lead to patent leakage or exposure of trade secrets in contexts such as research collaborations, enterprise engagements, or decentralized science (DeSci) platforms. Therefore, building a ``trustworthy yet structure-private'' verification mechanism has emerged as a new demand in generative scientific tasks~\cite{chen2024generative}.

To address the above challenges, this paper focuses on a task of high practical relevance: molecular toxicity repair. This task aims to systematically evaluate the ability of MLLMs to modify toxic small molecules into non-toxic or low-toxicity structures, serving as a key benchmark in drug discovery, toxicity prescreening, and the AI4S community~\cite{lin2025breaking}. Building upon our previous ToxiMol benchmark and its multi-dimensional evaluation chain ToxiEval, we further propose a structure-private verification framework—\textbf{ToxiEval-ZKP}—that introduces zero-knowledge proof (ZKP) mechanisms into the evaluation pipeline for the first time, enabling verifiability under conditions of structural invisibility.

Our core idea is to allow generative models to submit a ZKP-verified ``secure molecule commitment'', which proves that the molecule satisfies multi-dimensional evaluation criteria in the toxicity repair task (e.g., safety score, QED, SAS, etc.) without revealing its SMILES structure. For instance, a researcher may only declare: ``The generated molecule is non-toxic, with a median Lethal Dose (LD$_{50}$, the dose required to kill 50\% of test organisms) prediction above 800, and satisfies all drug-likeness constraints,'' along with a formal ZK proof. The verifier can validate the claim without accessing the specific molecular structure, thus preserving the privacy of the generated result while ensuring its scientific trustworthiness.

To achieve this goal, we design and implement a complete end-to-end ZK verification system, covering molecule-level metric checking logic at the circuit level, Poseidon-based structure commitment hashing, a \texttt{nullifier}-based deduplication and anti-replay mechanism, and evaluation pathways adapted to different task types (e.g., classification/regression). We conduct comprehensive testing of the system based on real-generated samples from MLLMs on ToxiMol tasks, validating the overall performance of ToxiEval-ZKP in terms of circuit efficiency, verification accuracy, and formal security.

The main contributions of this paper are as follows:
\begin{itemize}
  \item We propose a new paradigm for structure-private verification: This is the first work to introduce ZKP mechanisms into the molecular toxicity repair task, establishing a ``structure-invisible yet verifiable'' evaluation pipeline and offering a new direction for private evaluation in GenAI-based scientific~tasks.
  
  \item We design and implement a complete end-to-end ZK evaluation system: The system incorporates multiple toxicity-related evaluation criteria, circuit-level decision logic, structure hashing commitments, and nullifier tracking, supporting a variety of task types and evaluation strategies defined in the ToxiMol benchmark.
  
  \item We conduct multi-dimensional evaluation and analytical experiments: We validate the system’s functional correctness, scalability, and circuit complexity across multiple toxicity repair subtasks, demonstrating the broad applicability of our method in trustworthy evaluation for generative scientific~models.
\end{itemize}

\section{Preliminaries}

\subsection{Molecular Toxicity Repair Benchmark – ToxiMol}

ToxiMol~\cite{lin2025breaking} is a standardized benchmark designed to evaluate the generation and optimization capabilities of MLLMs in structure-level molecular toxicity repair tasks. In this task, the model takes a toxic molecule as input and, based on its Simplified Molecular Input Line Entry System (SMILES) representation, 2D structural image, and natural language description of the repair objective, is required to identify the potential toxicity mechanisms accurately, understand drug-likeness constraints, and generate structurally similar yet functionally detoxified molecules. To support this objective, ToxiMol is accompanied by a multi-dimensional evaluation framework—\textbf{ToxiEval}—which provides a comprehensive assessment of the generated molecules. The specific evaluation metrics and threshold definitions are provided in Table~\ref{tab:toxieval_metrics}.

\begin{table}[htbp]
  \small
  \renewcommand{\arraystretch}{1.25}
  \setlength{\tabcolsep}{5pt}
  \caption{Evaluation Metrics and Thresholds in ToxiEval~\cite{lin2025breaking}}
  \label{tab:toxieval_metrics}
  \rowcolors{2}{white}{gray!3} 
  \begin{tabularx}{\textwidth}{>{\columncolor{metricblue}}l|>{\raggedright\arraybackslash}X|c|c}
    \rowcolor{headergray}
    \textbf{Metric} & \textbf{Description} & \textbf{Range} & \textbf{Threshold for Success} \\
    \hline
    \textbf{Safety Score} & Indicates toxicity mitigation based on TxGemma~\cite{wang2025txgemma} classification result & 0–1 or binary & $=1$ (binary) or $>0.5$ (LD50 task) \\
    \textbf{QED} & Quantitative estimate of drug-likeness; higher is better & 0–1 & $\geq 0.5$ \\
    \textbf{SAS} & Synthetic accessibility; lower is better & 1–10 & $\leq 6$ \\
    \textbf{RO5} & Number of Lipinski’s rule violations (should be minimal) & Integer ($\geq 0$) & $\leq 1$ \\
    \textbf{SS} & Tanimoto similarity between original and repaired molecules & 0–1 & $\geq 0.4$ \\
  \end{tabularx}
\end{table}

Although the original purpose of ToxiEval-ZKP is to offer structural privacy protection and verifiable evaluation mechanisms for the ToxiMol benchmark, its underlying methodology is highly generalizable and extensible. In the current field of molecular generation, many studies have explored structure optimization and toxicity control using autoregressive architectures and diffusion models~\cite{wang2025txgemma, schneuing2024structure}. However, these methods often overlook critical issues related to structural security and verification credibility in data sharing, model evaluation, and collaborative research processes. We hope that this work will promote greater attention within the molecular generation community to the trustworthy assessment and privacy-compliant mechanisms, thereby extending the trust boundaries of existing research paradigms.

\subsection{ZKP}

ZKP is a class of cryptographic protocols that allow a prover to convince a verifier of the truth of a given statement without revealing any information about the underlying data itself. Its core properties include completeness, soundness, and zero-knowledge. Among them, the zero-knowledge property ensures that the verifier learns nothing beyond the validity of the statement being proven. Based on the interaction pattern, ZKPs can be classified into \textit{Interactive ZKPs} and \textit{Non-Interactive ZKPs}~\cite{sun2021survey}. In practical applications, non-interactive ZKPs are more widely adopted due to their higher communication efficiency and better compatibility with on-chain environments. Common implementations include zk-SNARKs~\cite{chen2022review}, zk-STARKs~\cite{panait2020using}. Among these, zk-SNARKs are particularly favored in blockchain, decentralized autonomous organizations (DAOs), and Web3.0 ecosystems due to their concise proofs and fast verification~\cite{sun2021survey}.

In recent years, ZKP has been extended to the machine learning domain, giving rise to an emerging research area known as \textit{Zero-Knowledge Proof Verifiable Machine Learning} (ZKP-VML). Under this framework, ZKP is used to prove the correctness of model inference or the integrity of input data without revealing underlying model parameters or sensitive data~\cite{peng2025survey}. In the GenAI field, ZKP has also begun to show promise; methods such as zkLLM~\cite{sun2024zkllm}, zkGPT~\cite{qu2025zkgpt}, explore incorporating ZKP mechanisms into the inference verification of LLMs, enabling provenance tracking and provable behavior for generated content. In the area of trustworthy visual content verification, works like ZK-IMG~\cite{kang2022zk} leverage zk-SNARKs to provide ZKPs of image editing processes, ensuring the authenticity and integrity of operations such as cropping and enhancement without disclosing the original images.

Unlike the above approaches, which primarily focus on constructing complex ZKP systems to verify the inference of deep models or language generation processes, our work does not rely on circuit modeling for large-scale model forward passes. Instead, we adopt a result-level verification perspective and design a ZKP construction paradigm specifically for scientific task outputs. Compared to the high construction complexity and computational overhead of ZKP for Machine Learning (ML)~/~LLM, our approach adopts a task-aware, lightweight proof path, showcasing the potential of ZKP in scientific generative tasks where structural sensitivity, privacy constraints, and verification requirements are particularly critical.

\section{Approach}

\subsection{Problem Definition}

For each repaired molecule $M$ generated by an MLLM, we define a vector of structural and toxicity evaluation metrics as $\mathbf{v} = \{v_{\text{valid}}, v_{\text{safe}}, v_{\text{QED}}, v_{\text{SAS}}, v_{\text{Lip}}, v_{\text{Sim}}\}$. Here, $v_{\text{valid}} \in \{0,1\}$ indicates whether the molecule is structurally valid, i.e., whether RDKit can successfully parse it into a valid molecular graph. This is the primary judgment criterion in the ToxiEval evaluation chain—if this item is 0, the remaining metrics will not be evaluated. The semantic definitions and evaluation methods of the other five metrics are provided in Table~\ref{tab:toxieval_metrics}. 

Based on this, we aim to construct a formal ZKP protocol, where the prover convinces the verifier of the following statement:

\[
\pi :\ \text{Prover convinces Verifier that } \forall i,\ v_i \geq \theta_i, \text{ without revealing any } v_i
\]

That is, without disclosing the specific values of any metric in $\mathbf{v}$, the prover only returns a Boolean result indicating whether all evaluation thresholds are satisfied, thereby enabling structure-private validation of molecular generation quality. The core challenge of this problem lies in efficiently mapping multiple floating-point or Boolean indicators into zk-SNARK-compatible statements through circuit construction, while ensuring verification efficiency and preventing the leakage of sensitive internal molecular properties.

\subsection{System Architecture}

We construct a modular zero-knowledge verification system, named \textbf{ToxiEval-ZKP}, to validate whether a generated molecule satisfies multiple structural and toxicity evaluation criteria without disclosing its structure or internal properties. The system consists of four core components: a data processor, a proof generator, a proof verifier, and a Circom verification circuit. The runtime procedure is outlined in Algorithm~\ref{alg:zkp_pipeline}.

\begin{algorithm}[!htbp]
\caption{ToxiEval-ZKP Runtime Pipeline}
\label{alg:zkp_pipeline}
\begin{algorithmic}
\REQUIRE Molecular evaluation results $R$; task type $t \in \{0,1\}$; threshold vector $\boldsymbol{\theta}$
\ENSURE Verification result $\text{verification\_result} \in \{0,1\}$; commitment $\text{commitment} \in \mathbb{F}_p$; nullifier $\text{nullifier} \in \mathbb{F}_p$

\STATE \textbf{Data Processing:}
\STATE Extract metrics: $\mathbf{v} = \{v_{\text{valid}}, v_{\text{safe}}, v_{\text{QED}}, v_{\text{SAS}}, v_{\text{Lip}}, v_{\text{Sim}}\}$ from $R$
\STATE Normalize values: scale to integer representation ($\times 10^6$)
\STATE Generate random salt $s \leftarrow \mathcal{U}(\mathbb{F}_p)$

\STATE \textbf{Circuit Verification:}
\FOR{each metric $v_i \in \mathbf{v}$}
  \STATE Verify constraint: $\text{check}_i \gets \mathsf{EvalCircuit}(v_i, \theta_i, t)$
\ENDFOR
\STATE $\text{verification\_result} \gets \bigwedge_{i} \text{check}_i$

\STATE \textbf{Commitment Generation:}
\STATE $\text{commitment} \gets \mathsf{Poseidon}(\mathbf{v}, s)$
\STATE $\text{nullifier} \gets \mathsf{Poseidon}(\text{commitment}, t)$

\STATE \textbf{Proof Generation \& Verification:}
\STATE Generate zk-SNARK proof: $\pi \gets \mathsf{Prove}(\mathbf{v}, s, t, \boldsymbol{\theta})$
\STATE Verify proof: $\text{valid} \gets \mathsf{Verify}(\pi, t, \boldsymbol{\theta})$
\IF{$\text{nullifier} \in \mathcal{N}$}
  \RETURN $\text{false}$ (replay attack detected)
\ENDIF
\STATE Add $\text{nullifier}$ to $\mathcal{N}$
\RETURN $\text{valid} \land (\text{verification\_result} = 1)$

\end{algorithmic}
\end{algorithm} 

\subsubsection{Data Processor}
This module extracts the structural metric vector $\mathbf{v} = \{v_{\text{valid}}, v_{\text{safe}}, v_{\text{QED}}, v_{\text{SAS}}, v_{\text{Lip}}, v_{\text{Sim}}\}$ from evaluation result files and transforms it into integer-form private inputs required by the Circom circuit. To support all 11 toxicity repair tasks~\cite{lin2025breaking} included in the ToxiMol dataset, we design a unified value normalization~framework:
\begin{itemize}
  \item For binary classification tasks (e.g., AMES, Carcinogens~\cite{lin2025breaking}), the \texttt{toxic} class is mapped to 0, and the \texttt{non-toxic} class is mapped to $10^6$.
  \item For continuous tasks (e.g., LD50), normalized scores are multiplied by $10^6$ and then rounded to~integers.
  \item All floating-point metrics ($v_{\text{QED}}, v_{\text{SAS}}, v_{\text{Sim}}$) use the same scaling factor to ensure six-digit decimal~precision.
\end{itemize}
Additionally, the data processor performs a set of pre-checks, including SMILES syntax validation, value range checks, and logical consistency verification. The system also includes default-value padding and fault-tolerant mechanisms to improve robustness and adaptability to diverse evaluation outputs.

\subsubsection{ZK Proof Generator}
This module constructs a zk-SNARK proof $\pi$ using Circom and SnarkJS. The private inputs include the metric vector $\mathbf{v}$ and a random salt $s$, while the public inputs consist of the task type $t$ and the corresponding threshold vector $\boldsymbol{\theta} = \{\theta_{\text{safe}}, \theta_{\text{QED}}, \theta_{\text{SAS}}, \theta_{\text{Lip}}, \theta_{\text{Sim}}\}$. Before proof generation, the system computes a structural commitment using the Poseidon hash function:
\[
\text{commitment} = \mathsf{Poseidon}(v_{\text{valid}},\ v_{\text{safe}},\ v_{\text{QED}},\ v_{\text{SAS}},\ v_{\text{Lip}},\ v_{\text{Sim}},\ s)
\]
This commitment serves as a unique structural summary of the repaired molecule, allowing for future consistency checks without exposing any individual $v_i$ values.

\subsubsection{Verifier}
The verifier receives three public outputs: the zk proof $\pi$, the structural commitment value $\text{commitment}$, and a \texttt{nullifier} for replay protection. Here, $\pi$ is the zk-SNARK proof generated by the prover, while both $\text{commitment}$ and $\text{nullifier}$ are computed internally by the Circom circuit using the Poseidon hash function. To prevent multiple submissions of the same repaired molecule, a nullifier mechanism is introduced:
\[
\text{nullifier} = \mathsf{Poseidon}(\text{commitment}, t)
\]
If a particular $\text{nullifier}$ is submitted more than once, the system will reject all subsequent requests, ensuring that each repaired molecule is verified only once. The verifier applies the general-purpose circuit validation function $\mathsf{Verify}(\pi, \text{public\_inputs})$ and outputs a Boolean value $\text{verification\_result} \in \{0,1\}$ indicating whether all structural and toxicity thresholds are satisfied.

\subsubsection{Verification Circuit}
At the core of ToxiEval-ZKP lies the Circom verification circuit, which performs complete validation of whether a repaired molecule satisfies task-specific multi-dimensional threshold constraints and outputs commitment data to support consistency verification and replay protection. The circuit takes as private inputs the vector $\mathbf{v} = \{v_{\text{valid}}, v_{\text{safe}}, v_{\text{QED}}, v_{\text{SAS}}, v_{\text{Lip}}, v_{\text{Sim}}\}$ and a random salt $s$, and as public inputs the task type $t \in \{0,1\}$ and the threshold set $\boldsymbol{\theta} = \{\theta_{\text{safe}}, \theta_{\text{QED}}, \theta_{\text{SAS}}, \theta_{\text{Lip}}, \theta_{\text{Sim}}\}$. The core verification logic is defined as:

\[
\mathsf{EvalCircuit}(\mathbf{v};\ \boldsymbol{\theta},\ t) =
\begin{cases}
1, & \text{if } \forall i,\ v_i \geq \theta_i \text{ (evaluated under task type } t\text{)} \\
0, & \text{otherwise}
\end{cases}
\]

Here, the task type $t$ determines how the safety metric $v_{\text{safe}}$ is evaluated: for binary classification tasks ($t = 0$), $v_{\text{safe}}$ must be strictly equal to $10^6$; for regression tasks ($t = 1$, e.g., LD50), the value must be greater than or equal to the predefined threshold $\theta_{\text{safe}}$. The circuit uses Boolean comparators to constrain each $v_i$ against its corresponding $\theta_i$, and combines all results using an AND gate to produce a single Boolean signal $verification\_result$, which returns 1 only when all metrics meet the required thresholds.

Additionally, the circuit internally implements the generation logic for both the structural commitment and the replay-prevention nullifier. Specifically, the Poseidon hash function is used to salt-hash the structural vector and salt, yielding:
\begin{align*}
\text{commitment} &= \mathsf{Poseidon}(\mathbf{v},\ s) \\
\text{nullifier}  &= \mathsf{Poseidon}(\text{commitment},\ t)
\end{align*}
These outputs support subsequent off-chain or on-chain consistency checks, preventing duplicate submissions and ensuring the system remains auditable while preserving privacy.
Finally, the circuit outputs three results: the verification flag $\text{verification\_result} \in \{0,1\}$, the structure commitment $\text{commitment} \in \mathbb{F}_p$, and the nullifier $\text{nullifier} \in \mathbb{F}_p$, where $\mathbb{F}_p$ denotes the finite field over which the zk hash function operates. Together, these three outputs constitute the complete ZKP interface of the ToxiEval-ZKP system.

\section{Experiments}

To comprehensively evaluate the performance, scalability, and security of the ToxiEval-ZKP system, we design three core experiments: system scalability testing, formal verification of security properties, and analysis of Circom circuit complexity. All experiments are conducted on an Apple M3 processor with 48GB of memory. The experimental data are based on repaired molecules generated by Claude-3.7 Sonnet on the ToxiMol benchmark~\cite{lin2025breaking}. After evaluation by the ToxiEval module, a total of 1,687 molecule samples were obtained, including 377 successful repairs, 1,151 failed repairs, and 159 invalid SMILES.

The ToxiEval-ZKP system is implemented using the Groth16 ZKP protocol, employing the BN128 elliptic curve and the Poseidon hash function. The constructed Circom verification circuit consists of 1,774 constraints (994 linear and 780 nonlinear), with a circuit depth of 8 layers, supporting 6 public inputs, 7 private inputs, and 3 public outputs.

\subsection{Scalability Testing}

We evaluate the system's performance at different input scales: 10, 50, 100, 200, and 377 molecules. Each configuration is run three times, and we record the total processing time, average processing time per molecule, throughput, peak memory usage, and verification success rate.

As shown in Table~\ref{tab:scalability_results}, ToxiEval-ZKP demonstrates excellent linear scalability. At the largest scale (377 molecules), the average total processing time is 36.95 seconds; the per-molecule processing time remains steady at 0.098 seconds, and throughput reaches 10.20 molecules per second. The verification success rate is 100\% across all test scales, indicating high system reliability. Memory consumption grows linearly with input size, from 727MB (10 molecules) to 981MB (377 molecules), demonstrating efficient memory usage. Further standard deviation analysis shows that the coefficient of variation for all metrics is less than 5\%, indicating very low-performance fluctuation.

\begin{table}[htbp]
\centering
\caption{Scalability Test Results of ToxiEval-ZKP System}
\label{tab:scalability_results}
\begin{tabular*}{\linewidth}{@{\extracolsep{\fill}}cccccc@{}}
\toprule
Molecules & Avg Total & Time per & Throughput & Peak Memory & Success \\
 & Time (s) & Molecule (s) & (mol/s) & (MB) & Rate (\%) \\
\midrule
10  & 1.056 ± 0.084 & 0.1056 ± 0.0084 & 9.53 ± 0.72 & 727.2 ± 39.1 & 100.0 \\
50  & 4.850 ± 0.025 & 0.0970 ± 0.0005 & 10.31 ± 0.05 & 791.6 ± 8.7 & 100.0 \\
100 & 9.762 ± 0.046 & 0.0976 ± 0.0005 & 10.24 ± 0.05 & 830.6 ± 10.1 & 100.0 \\
200 & 19.585 ± 0.042 & 0.0979 ± 0.0002 & 10.21 ± 0.02 & 880.9 ± 16.2 & 100.0 \\
377 & 36.947 ± 0.065 & 0.0980 ± 0.0002 & 10.20 ± 0.02 & 981.4 ± 32.0 & 100.0 \\
\bottomrule
\end{tabular*}
\end{table}

\begin{figure}[t]
  \centering
  \includegraphics[width=1\linewidth]{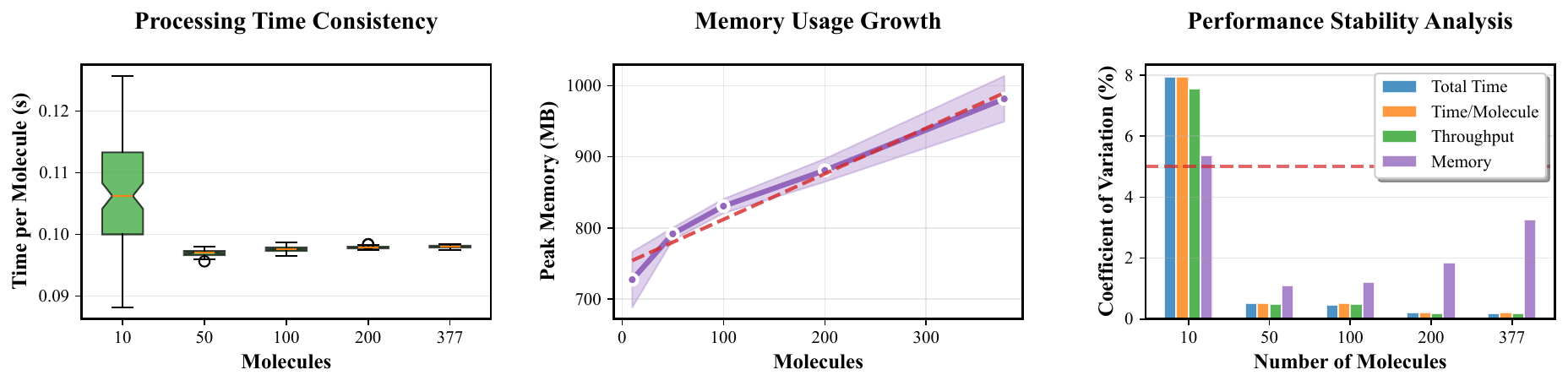}
  \caption{Scalability analysis of the ToxiEval-ZKP system across three dimensions. 
In the Memory Usage Growth panel, the red dashed line denotes the linear regression trend of memory usage with respect to the number of molecules, serving as a reference for the consistency between theoretical scalability expectations and observed behavior. 
In the Performance Stability Analysis panel, the red dashed line indicates the 5\% coefficient-of-variation stability threshold, below which the system performance can be regarded as consistent and reliable.}
  \label{fig:scalability_analysis}
\end{figure}

Figure~\ref{fig:scalability_analysis} illustrates the system's scalability across three dimensions. The processing time consistency analysis shows that per-molecule processing time remains stable with a standard deviation of $<$ 0.001s. Memory usage grows linearly from 727MB to 981MB, indicating efficient resource utilization. Performance stability analysis reveals all metrics maintain a coefficient of variation below 5\%, demonstrating robust reliability. These results confirm the system's linear scalability and suitability for large-scale molecular repair tasks.

\subsection{Formal Security Verification}

From a cryptographic perspective, we formally verify four core security properties of the ToxiEval-ZKP system. First, \textit{completeness} requires that the system correctly accepts all valid molecule inputs that meet the verification criteria—i.e., all legally generated repaired molecules should yield valid proofs. Second, \textit{soundness} demands that the system must reject all non-conforming inputs, ensuring that invalid repairs cannot pass verification. Third, \textit{zero-knowledge} guarantees that the proof process does not leak any private molecular structural information, maintaining information-theoretic privacy. Lastly, \textit{attack resistance} requires the system to remain robust and correctly identify malicious inputs under various adversarial scenarios, including boundary inputs, type confusion, overflow injection, and replay attacks.

The tests are conducted as follows: 50 successfully repaired molecules are used to assess proof acceptance capability (completeness); 60 invalid inputs (30 invalid SMILES and 30 molecules not meeting the criteria) are used to assess rejection capability (soundness); 20 proof samples are used for entropy analysis and leakage detection to evaluate zero-knowledge; and four types of attack scenarios are designed to test the system's robustness.

As shown in Table~\ref{tab:security_results}, the system performs well across all tests. In the completeness test, all 50 valid molecules passed verification (100\%); in the soundness test, all 60 invalid samples were correctly rejected by the system. The zero-knowledge test detected no information leakage, with an average entropy of 0.415, indicating strong randomness in structural commitments. Under all four attack scenarios, the system remained stable; in particular, the nullifier mechanism effectively detected and blocked repeated submissions during replay attacks.

\begin{table}[htbp]
\centering
\caption{Security Formal Verification Results of ToxiEval-ZKP System}
\label{tab:security_results}
\begin{tabular*}{\linewidth}{@{\extracolsep{\fill}}lcccl@{}}
\toprule
Security Property & Test & Passed & Success & Key Metrics \\
 & Samples & Tests & Rate (\%) & \\
\midrule
Completeness & 50 & 50 & 100.0 & All valid proofs accepted \\
Soundness & 60 & 60 & 100.0 & All invalid inputs rejected \\
Zero-Knowledge & 20 & 20 & 100.0 & Avg entropy: 0.415, no leakage \\
Attack Resistance & 4 & 4 & 100.0 & Replay detection effective \\
\midrule
\textbf{Overall Security Score} & \textbf{134} & \textbf{134} & \textbf{100.0} & \textbf{Perfect security performance} \\
\bottomrule
\end{tabular*}
\end{table}

\subsection{Circuit Complexity Analysis}

We conduct a detailed analysis of the Circom verification circuit from four dimensions: constraint density, critical path depth, parallelization potential, and memory access pattern. The constraint density analysis evaluates the distribution of computational complexity across circuit components, while the critical path analysis identifies performance bottlenecks. The parallelization analysis assesses opportunities for parallel optimization, and the memory pattern analysis evaluates memory efficiency.

\begin{figure}[t]
  \centering
  \includegraphics[width=1\linewidth]{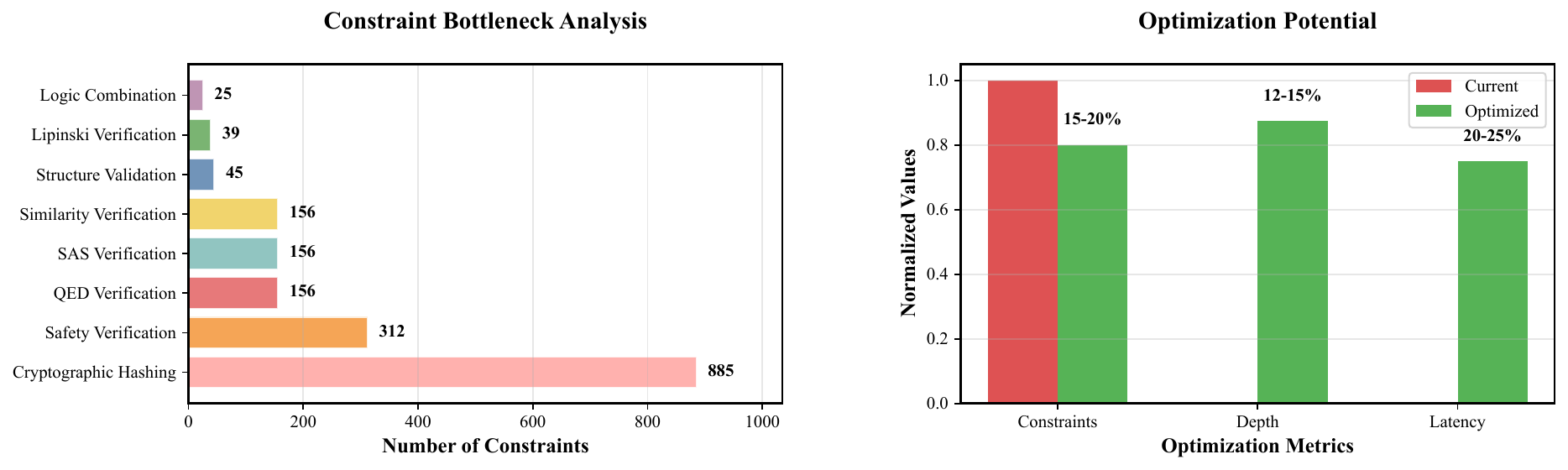}
  \caption{Circuit complexity analysis of the ToxiEval-ZKP system.} 
  \label{fig:constraint_distribution}
\end{figure}

As shown in Figure~\ref{fig:constraint_distribution}, the constraint bottleneck analysis reveals that cryptographic hash operations account for 49.9\% of the constraints (885 constraints), making them the primary source of system complexity. Safety verification accounts for 17.6\% (312 constraints), while the remaining evaluation standards (QED, SAS, similarity) each account for approximately 8.8\% (156 constraints). This distribution clearly identifies cryptographic hashing as the dominant computational bottleneck, followed by safety verification processes.

The optimization potential analysis demonstrates substantial improvement opportunities across three key metrics: constraint optimization (15--20\% reduction potential), depth optimization (12--15\%), and latency optimization (20--25\%). These findings suggest that targeted optimizations, focusing on simplifying hash functions, merging constraints, eliminating redundancy, and restructuring circuits, can significantly enhance system performance while maintaining security guarantees.

\section{Discussion}

In future research, we plan to extend the current zero-knowledge verification paradigm from structure-based molecular repair tasks to more complex domains involving generative models. Diffusion models have achieved remarkable progress in molecular image generation and structure completion tasks. However, due to the high complexity of their generation processes and the opacity of their latent spaces, verifying whether the generated results satisfy scientific constraints—such as physical feasibility, biological activity, or synthetic accessibility—remains a significant challenge.

\section{Conclusion}

This paper presents \textbf{ToxiEval-ZKP}, the first system to realize a structure-private verification mechanism for molecular toxicity repair tasks. By integrating ZKP protocols with a multi-metric evaluation chain, the system strikes a balance between verifiability and privacy protection, demonstrating strong scalability, security, and efficiency across multiple evaluation dimensions. In future work, we plan to further extend this mechanism to molecular generation tasks driven by LLMs and diffusion models.

\bibliographystyle{unsrtnat}
\bibliography{references}

\end{document}